\documentclass[12pt]{iopart}

\usepackage{iopams}
\usepackage{graphicx}
\usepackage{dcolumn}
\usepackage{bm}

\begin{document}

\title{Diluted antiferromagnet in a ferromagnetic enviroment}
\author{M O Hase and J F F Mendes}
\address{Departamento de F\'isica da Universidade de Aveiro, 3810-193 Aveiro, Portugal}
\ead{mhase@if.usp.br}

\begin{abstract}
The question of robustness of a network under random ``attacks'' is treated in the framework of critical phenomena. The persistence of spontaneous magnetization of a ferromagnetic system to the random inclusion of antiferromagnetic interactions is investigated. After examing the static properties of the quenched version (in respect to the random antiferromagnetic interactions) of the model, the persistence of the magnetization is analysed also in the annealed approximation, and the difference in the results are discussed.
\end{abstract}

\maketitle


\section{Introduction}

The investigation on resilience of networks, which examines the persistence/breakdown of some global properties of a graph under, for instance, removal of vertices or edges, is known to have practical importance. Many ``real networks'' (Internet, highways, many biological systems, \textit{et cetera}) depend on the fact that there exist links between the nodes to ensure their functionality. A damage that breaks the interconnection between vertices can trigger a profound impact on the network\cite{AJB00}. The connection of these questions to the percolation theory is clear, which has been a powerful tool to tackle these problems\cite{CEbAH00} (see also \cite{DM01, CNSW00}).

In this work, the question of resilience of networks is analysed in the framework of critical phenomena of magnetic systems, and is not directly related to the classical problems concerning the existence of giant components in a graph\cite{B85, MR95}.

Critical phenomena have been exhaustively exploited in networks\cite{DGM07, DM03}, and it was immediately noticed that it displays a quite rich thermodynamical behaviour when compared to regular lattice structures\cite{G00, BW00, NCCSHW04, LVVZ02} or gives place to nontrivial effects\cite{BCD05, KDM07}. This shows the role of the topology of the underlying graph on critical behaviour. Moreover, networks constitute a path toward a more ``realistic'' system of mean - field models, the main cause being the introduction of (finite) connectivity as a parameter\cite{VB85, KS87, MP87, MdD87, dAdDM88, MP01}.

This work will focus on the persistence of a order imposed by a background enviroment to random inclusion of opposite effects on the network. More precisely, starting from a ferromagnetic system, some of its interactions are weakened (or even changed) by antiferromagnetic ones. The criterion for the ``functionality'' of the network will be the ferromagnetic order of the model, and by ``resilience'' it means how strong is the spontaneous magnetization against random introduction of antiferromagnetic interaction that contributes to the disorder. The model is detailed in section II, and the thermodynamical analysis of its quenched version is given in section III, followed by a characterization of the order - disorder critical line in the annealed approximation in section IV. The critical line that determines the breakdown of the ordered phase is discussed in section V, and the last section is devoted to conclusions.


\section{Model}

Consider a (finite) graph $\Lambda_{N}$ ($|\Lambda_{N}|=N$) where each vertex $x\in\Lambda_{N}$ allocates an Ising spin $\sigma_{x}\in\{-1,1\}$. The Hamiltonian is given by
\begin{eqnarray}
H(\{\sigma_{x}\}, \{a_{xy}\}) = -\frac{J}{2N}\left(\sum_{x\in\Lambda_{N}}\sigma_{x}\right)^{2} + J_{A}\sum_{\stackrel{x,y\in\Lambda_{N}}{x<y}} a_{xy}\sigma_{x}\sigma_{y}\,,
\label{Hamilton}
\end{eqnarray}
where the first term represents the ferromagnetic interaction ($J/N>0$) between every pair of spins and the second sum is the antiferromagnetic interaction ($-J_{A}<0$) between some pair of spins. Therefore, the network is a complete graph where the edges represent interactions. Note that the antiferromagnetic interactions do not constitute new edges on the graph themselves, but they weaken the pre - existent ferromagnetic edges competitively. The matrix $\{a_{xy}\}$ decides if two vertices, $x$ and $y$, have antiferromagnetic interaction ($a_{xy}=1$) or not ($a_{xy}=0$), and the couplings $\{a_{xy}\}$ are independent and identically distributed random variables that obey the distribution
\begin{eqnarray}
\mathcal{P}(a_{xy}) = \left(1-\frac{p}{N}\right)\delta_{a_{xy},0} + \frac{p}{N}\delta_{a_{xy},1}\,.
\label{Pa}
\end{eqnarray}

It is known that the quenched version of model (\ref{Hamilton}) without the ferromagnetic term (the first sum) is a frustrated system with a spin glass phase at low temperatures\cite{HdAS06}(since the model has a single sublattice). Therefore, between the two effects of antiferromagnetic interactions, namely the contribution to disorder and a contribution to ferromagnetic order (due to the effect of frustration arisen in some loops of the graph), the former dominates over the later.

As usual, the partition function is
\begin{eqnarray}
Z(\beta,\{a_{xy}\}) := \mbox{Tr}_{\{\sigma_{x}\}}e^{-\beta H(\{\sigma_{x}\},\{a_{xy}\})}\,,
\label{Z}
\end{eqnarray}
where $\beta$ is the inverse of the temperature $T$ ($k_{B}=1$) and the trace indicates the sum over $2^{N}$ states. This partition function also depends on the configuration of the antiferromagnetic interactions. Finally, given a function $g=g(\{a_{xy}\})$ of the set of random variables $\{a_{xy}\}$, the average of $g$ over the configuration $\{a_{xy}\}$ will be denoted as
\begin{eqnarray}
\langle g\rangle := \int g(\{a_{xy}\}) \prod_{x<y}\mathcal{P}(\{a_{xy}\})da_{xy}\,.
\label{<>}
\end{eqnarray}


\section{Thermodynamics}

The (quenched) free energy $f_{q}$ of the model is evaluated through the replica trick
\begin{eqnarray}
f_{q}(\beta) = -\lim_{N\rightarrow\infty}\frac{1}{\beta N}\langle\ln Z(\beta)\rangle = -\lim_{n\rightarrow 0}\lim_{N\rightarrow\infty}\frac{1}{\beta Nn}\ln\langle Z^{n}(\beta)\rangle\,,
\label{<lnz>}
\end{eqnarray}
where besides the analytic continuation in $n\in\mathbb{N}\rightarrow n\in\mathbb{R}$, the order of the limits $N\rightarrow\infty$ and $n\rightarrow 0$ was changed as usual. Then, one should compute $\langle Z^{n}(\beta)\rangle$, which can be casted as
\begin{eqnarray}
\fl\nonumber \langle Z^{n}(\beta)\rangle = \left(\frac{N\beta J}{2\pi}\right)^{\frac{n}{2}}\left(\prod_{\alpha=1}^{n}\int\limits_{\mathbb{R}}d\lambda^{\alpha}\right)\;\mbox{Tr}_{\{\sigma_{x}^{\alpha}\}}\exp\Bigg[ -\frac{Np}{2} - \frac{N\beta J}{2}\sum_{\alpha=1}^{n}\left(\lambda^{\alpha}\right)^{2} + \\
 + \beta J\sum_{\alpha=1}^{n}\lambda^{\alpha}\sum_{x\in\Lambda_{N}}\sigma_{x}^{\alpha} + \frac{p}{2N}\sum_{x,y\in\Lambda_{N}}e^{-\beta J_{A}\sum_{\alpha=1}^{n}\sigma_{x}^{\alpha}\sigma_{y}^{\alpha}} + \mathcal{O}(1)\Bigg]\,.
\label{zq1}
\end{eqnarray}
Throughout this work, upper and lower indices at spin variables indicate replica index (greek letter) and site position (roman letter), respectively.

Introducing the order parameter\cite{M98}
\begin{eqnarray}
\psi(\mu) := \frac{1}{N}\sum_{x\in\Lambda_{N}}\delta_{\{\sigma_{x}^{\alpha}\},\{\mu^{\alpha}\}}\,,
\label{op}
\end{eqnarray}
where $\delta_{\{\sigma_{x}^{\alpha}\},\{\mu^{\alpha}\}}:=\prod_{\alpha=1}^{n}\delta_{\sigma_{x}^{\alpha},\mu^{\alpha}}$, the equation (\ref{zq1}) can be written as
\begin{eqnarray}
\langle Z^{n}(\beta)\rangle \sim \left(\frac{N\beta J}{2\pi}\right)^{\frac{n}{2}}\left(\prod_{\alpha=1}^{n}d\lambda^{\alpha}\right)\int\mathcal{D}\psi\mathcal{D}\hat\psi e^{-N\phi_{q}[\psi,\hat\psi](\beta,\{\lambda^{\alpha}\})}\,,
\label{zq2}
\end{eqnarray}
where
\begin{eqnarray}
\fl\nonumber \phi_{q}[\psi,\hat\psi](\beta,\{\lambda^{\alpha}\}) := \frac{p}{2} + \frac{\beta J}{2}\sum_{\alpha=1}^{n}\left(\lambda^{\alpha}\right)^{2} + \mbox{Tr}_{\{\mu^{\alpha}\}}\psi(\mu)\hat\psi(\mu) - \\
 - \frac{p}{2}\;\mbox{Tr}_{\{\mu^{\alpha}\}}\;\mbox{Tr}_{\{\tau^{\alpha}\}}\psi(\mu)\psi(\tau)e^{-\beta J_{A}\sum_{\alpha=1}^{n}\mu^{\alpha}\tau^{\alpha}} - \ln\zeta[\hat\psi](\beta,\{\lambda^{\alpha}\})
\label{phiq}
\end{eqnarray}
and
\begin{eqnarray}
\zeta[\hat\psi](\beta,\{\lambda^{\alpha}\}) := \mbox{Tr}_{\{\sigma^{\alpha}\}}\exp\left[\beta J\sum_{\alpha=1}^{n}\lambda^{\alpha}\sigma^{\alpha} + \hat\psi(\sigma)\right]\,.
\label{zeta}
\end{eqnarray}

Apart from a factor $\beta$, $\phi_{q}[\psi,\hat\psi](\beta,\{\lambda^{\alpha}\})$ is just the variational free energy. The equation (\ref{zq2}) suggests that one should invoke the saddle - point method to determine the stationary free energy. The extremum conditions, necessary to ensure the infimum of $\phi_{q}$ over the suitable functions ($\psi$ and $\hat\psi$) and variables ($\{\lambda^{\alpha}\}$), are
\begin{eqnarray}
\left\{
\begin{array}{lcl}
\psi(\mu) & = & \displaystyle\frac{1}{\zeta[\hat\psi](\beta,\{\lambda^{\alpha}\})} \exp\left[\beta J\sum_{\alpha=1}^{n}\lambda^{\alpha}\mu^{\alpha} + \hat\psi(\mu)\right] \\
 & & \\
\hat\psi(\mu) & = & p\;\mbox{Tr}_{\{\tau^{\alpha}\}}\psi(\tau)\exp\displaystyle\left[-\beta J_{A}\sum_{\alpha=1}^{n}\mu^{\alpha}\tau^{\alpha}\right] \\
 & & \\
\lambda^{\alpha} & = & \displaystyle\frac{1}{\zeta[\hat\psi](\beta,\{\lambda^{\alpha}\})} \;\mbox{Tr}_{\{\sigma^{\alpha}\}}\sigma^{\alpha}\exp\left[\beta J\sum_{\alpha=1}^{n}\lambda^{\alpha}\sigma^{\alpha} + \hat\psi(\sigma)\right]
\end{array}
\right.\,.
\label{extr}
\end{eqnarray}

In an attempt to solve the above equations, one should cast the replica symmetric \textit{Ansatz}
\begin{eqnarray}
\left\{
\begin{array}{lcl}
\psi(\mu)        & = & \psi\left(\sum_{\alpha=1}^{n}\mu^{\alpha}\right) = \displaystyle\int\limits_{\mathbb{R}}dh P(h)\frac{e^{\beta h\sum_{\alpha=1}^{n}\mu^{\alpha}}}{\left[2\cosh(\beta h)\right]^{n}} \\
 & & \\
\hat\psi(\mu)    & = & \hat\psi\left(\sum_{\alpha=1}^{n}\mu^{\alpha}\right) = p\displaystyle\int\limits_{\mathbb{R}}dy Q(y)\frac{e^{\beta y\sum_{\alpha=1}^{n}\mu^{\alpha}}}{\left[2\cosh(\beta y)\right]^{n}} \\
 & & \\
\lambda^{\alpha} & = & \lambda\,, \quad \forall\alpha\in\{1,\cdots,n\}
\end{array}
\right.\,,
\label{rs}
\end{eqnarray}
where $P$ and $Q$ are probability distributions.

In the $n\rightarrow 0$ limit, one can show that
\begin{eqnarray}
\fl\left\{
\begin{array}{lcl}
\zeta[\hat\psi](\beta,\{\lambda^{\alpha}\}) & = & e^{p} \\
\psi(x) & = & e^{-p}\displaystyle\sum_{r=0}^{\infty}\frac{p^{r}}{r!}\left(\prod_{j=1}^{r}\int\limits_{\mathbb{R}}dy_{j}Q(y_{j})\right)\exp\left[\beta x\left(J\lambda + \sum_{k=1}^{r}y_{k}\right)\right] \\
\hat\psi(x) & = & p\displaystyle\int\limits_{\mathbb{R}}dh P(h)\left[\frac{\cosh\left(\beta J_{A}-\beta h\right)}{\cosh\left(\beta J_{A}+\beta h\right)}\right]^{\frac{x}{2}} \\
\lambda & = & e^{-p}\displaystyle\sum_{r=0}^{\infty}\frac{p^{r}}{r!}\left(\prod_{j=1}^{r}\int\limits_{\mathbb{R}}dy_{j}Q(y_{j})\right)\tanh\left[\beta\left(J\lambda+\sum_{k=1}^{r}y_{k}\right)\right]
\end{array}
\right.\,,
\label{zphpl}
\end{eqnarray}
the (quenched) free energy $f_{q}$ is evaluated as
\begin{eqnarray}
\fl\nonumber \beta f(\beta) = \frac{\beta J\lambda^{2}}{2} + \frac{p}{2}\ln 2 + p\int\limits_{\mathbb{R}}dh P(h)\int\limits_{\mathbb{R}}dy Q(y)\ln\cosh\left[\beta\left(h+y\right)\right] - \\
\fl\nonumber - \frac{p}{2}\int\limits_{\mathbb{R}}dh_{1} P(h_{1})\int\limits_{\mathbb{R}}dh_{2} P(h_{2})\ln\Big\{ e^{\beta h_{1}}\cosh\left[\beta\left(h_{2}-J_{A}\right)\right] + e^{-\beta h_{1}}\cosh\left[\beta\left(h_{2}+J_{A}\right)\right] \Big\} - \\
\fl -\ln 2 - e^{-p}\sum_{r=0}^{\infty}\frac{p^{r}}{r!}\left[\prod_{s=1}^{r}\int\limits_{\mathbb{R}}dy_{s}Q(y_{s})\right]\ln\cosh\left[\beta\left(J\lambda + \sum_{j=1}^{r}y_{j}\right)\right]\,,
\label{fq}
\end{eqnarray}
the extremum condition leads $P$ and $Q$ to satisfy
\begin{eqnarray}
\left\{
\begin{array}{lcl}
P(h) & = & e^{-p}\displaystyle\sum_{r=0}^{\infty}\frac{p^{r}}{r!}\left(\prod_{j=1}^{r}\int\limits_{\mathbb{R}}dy_{j}Q(y_{j})\right)\delta\left(h - \left[J\lambda + \sum_{k=1}^{r}y_{k}\right]\right) \\
 & & \\
Q(y) & = & \displaystyle\int\limits_{\mathbb{R}}dh P(h)\delta\left( y + \frac{1}{\beta}\tanh^{-1}\left[\tanh(\beta J_{A})\tanh(\beta h)\right] \right)
\end{array}
\right.\,,
\label{PQ}
\end{eqnarray}
and $\lambda$ is calculated through the last equation of (\ref{zphpl}). The above set of equations (\ref{PQ}) can be unified as
\begin{eqnarray}
\fl\nonumber P(h) = e^{-p}\displaystyle\sum_{r=0}^{\infty}\frac{p^{r}}{r!}\left(\prod_{j=1}^{r}\int\limits_{\mathbb{R}}dh_{j}P(h_{j})\right)\times \\
\times\delta\left(h - \left[J\lambda - \frac{1}{\beta}\sum_{k=1}^{r}\tanh^{-1}\left[\tanh(\beta J_{A})\tanh(\beta h_{k})\right]\right]\right)\,.
\label{P}
\end{eqnarray}

In the context of replica method, the formulas for the magnetization, $m$, and the spin - glass order parameter, $q$, are given by
\begin{eqnarray}
m = \lim_{N\rightarrow\infty}\frac{1}{N}\sum_{x\in\Lambda_{N}}\lim_{n\rightarrow 0}\frac{1}{n}\sum_{\alpha=1}^{n}\mbox{Tr}_{\{\sigma_{z}^{\eta}\}}\sigma_{x}^{\alpha} \left\langle \prod_{\gamma=1}^{n}e^{-\beta H(\{\sigma_{u}^{\gamma}\},\{a_{uv}\})} \right\rangle
\label{mreplica}
\end{eqnarray}
and
\begin{eqnarray}
q = \lim_{N\rightarrow\infty}\frac{1}{N}\sum_{x\in\Lambda_{N}}\lim_{n\rightarrow 0}\frac{1}{n\left(n-1\right)}\sum_{\alpha\neq\theta}^{n}\mbox{Tr}_{\{\sigma_{z}^{\eta}\}}\sigma_{x}^{\alpha}\sigma_{x}^{\theta} \left\langle \prod_{\gamma=1}^{n}e^{-\beta H(\{\sigma_{u}^{\gamma}\},\{a_{uv}\})} \right\rangle\,.
\label{qreplica}
\end{eqnarray}

Using (\ref{PQ}), (\ref{mreplica}) and (\ref{qreplica}), it is possible to see that $\lambda=m$,
\begin{eqnarray}
m = \int\limits_{\mathbb{R}}dh P(h)\tanh(\beta h)\quad\textrm{and}\quad q = \int\limits_{\mathbb{R}}dh P(h)\tanh^{2}(\beta h)\,,
\label{mq}
\end{eqnarray}
as usual. The critical line is determined in the neighborhood of $m\sim 0$ and $q\sim 0$. In this regime, the field $h$ is expected to be narrowly distributed around $h=0$. Then, the \textit{Ansatz}
\begin{eqnarray}
\overline{\epsilon^{k}} := \int\limits_{\mathbb{R}}dh P(h)h^{k} = \mathcal{O}(\epsilon^{k})\,,\quad |\epsilon|\ll 1\,,
\label{ans}
\end{eqnarray}
is introduced in the equation (\ref{P}) to evaluate the transition lines\cite{NCCSHW04}. This \textit{Ansatz} leads to $m=\mathcal{O}(\epsilon)$ and $q=\mathcal{O}(\epsilon^{2})$, which means that the line of transition from ferromagnetic to disordered phase is governed by $\mathcal{O}(\epsilon)$ in the equation (\ref{P}) and the transition from paramagnetic to spin - glass phase by the order $\mathcal{O}(\epsilon^{2})$ ($\overline{\epsilon}=0$ is assumed in this case). As a result of these calculations, one has
\begin{eqnarray}
\beta_{c}J = 1 + p\tanh(\beta_{c}J_{A}) \quad \textrm{(order - disorder transition line)}\,,
\label{tcq}
\end{eqnarray}
and the transition line from spin - glass to paramagnetic phase is evaluated as
\begin{eqnarray}
\beta_{c}J = \frac{1}{2\left(J_{A}/J\right)}\ln\left(\frac{\sqrt{p}+1}{\sqrt{p}-1}\right)\,,\quad p>1\,.
\label{SG-P}
\end{eqnarray}

The phase diagram of the model, generated by the equations (\ref{tcq}) and (\ref{SG-P}), is presented in FIG 1. The ``$m\neq 0$'' phase displays spontaneous magnetization, and it may be a combination of a ferromagnetic and a mixed phases. The exact scenario of the ``$m\neq 0$'' phase can be established from a stability analysis\cite{dAT78}\footnote{To be more precise, the stability analysis can also change the exact location of the line that separates the spin - glass phase and the paramagnetic phase; however, the phase diagram showed in FIG 1 is believed to be qualitatively correct.}, which will not be provided here, since hereafter this work will focus on the critical line that separates the ordered phase and disordered one in comparison with the annealed version of the model.

\begin{figure}[htb]
\centering
\includegraphics[width=80.0mm, height=80.0mm]{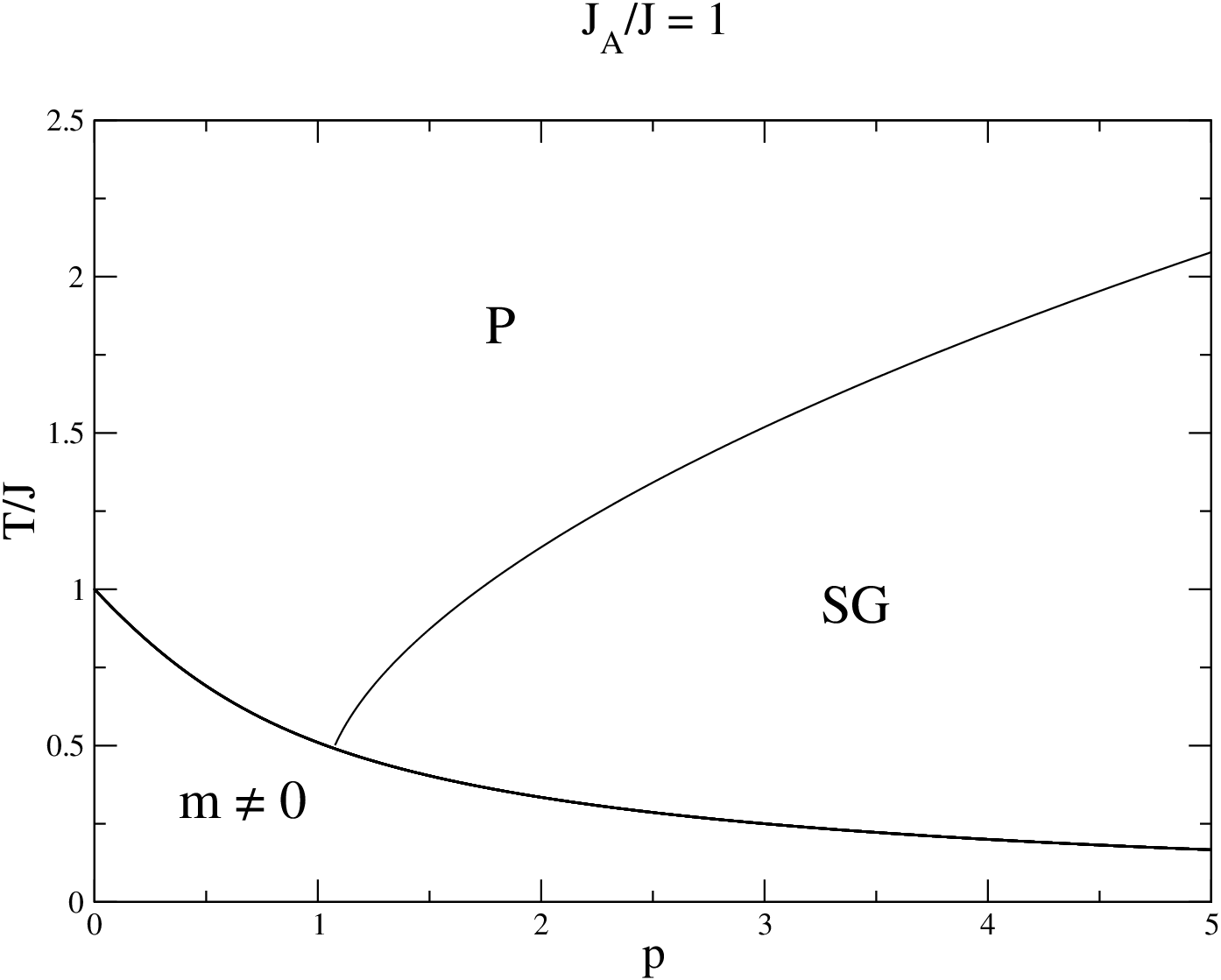}
\bigskip
\caption{Phase diagram with $J_{A}/J=1$.}
\label{1}
\end{figure}


\section{Annealed average}

In the annealed approximation, the free energy $f_{a}$ is written as
\begin{eqnarray}
f_{a}(\beta) = -\lim_{N\rightarrow\infty}\frac{1}{\beta N}\ln\langle Z(\beta)\rangle\,.
\label{f_an1}
\end{eqnarray}
The problem can be solved in the annealed approximation in a simpler way. Although the interesting case is the quenched one (which was provided in the previous section), the results will be derived for comparison.

It is straighforward to show that
\begin{eqnarray}
\fl\nonumber \langle Z(\beta)\rangle = \mbox{Tr}_{\{\sigma_{x}\}}\exp\Bigg\{ -\frac{Np}{2} + \frac{Np}{2}\cosh(\beta J_{A}) + \frac{1}{2N}\Big[\beta J - p\sinh(\beta J_{A})\Big]\Big(\sum_{x\in\Lambda_{N}}\sigma_{x}\Big)^{2} + \\
 + \mathcal{O}(1) \Bigg\}\,,
\label{<z>}
\end{eqnarray}
and introducing the order parameter
\begin{eqnarray}
m_{N} := \frac{1}{N}\sum_{x\in\Lambda_{N}}\sigma_{x}\,,
\label{m}
\end{eqnarray}
it is easy to show that
\begin{eqnarray}
\langle Z(\beta)\rangle \sim \int\limits_{\mathbb{R}}dm_{N} e^{-N\phi_{a}(\beta,m_{N})}
\label{<z>2}
\end{eqnarray}
for sufficiently large $N$, where
\begin{eqnarray}
\fl\nonumber \phi_{a}(\beta,m_{N}) := \frac{p}{2} - \frac{p\cosh(\beta J_{A})}{2} - \frac{\beta J}{2}m_{N}^{2} + \frac{p\sinh(\beta J_{A})}{2}m_{N}^{2} + \frac{m_{N}}{2}\ln\left(\frac{1+m_{N}}{1-m_{N}}\right) + \\
 + \frac{1}{2}\ln\left(1-m_{N}^{2}\right) - \ln 2\,.
\label{phia}
\end{eqnarray}

Therefore, in the thermodynamic limit, with $m:=\lim_{N\rightarrow\infty}m_{N}$, the free energy can be written as
\begin{eqnarray}
f_{a}(\beta) = \frac{1}{\beta}\inf_{m}\left\{\phi_{a}(\beta,m)\right\}\,,
\label{f_an2}
\end{eqnarray}
where the infimum of $\phi_{a}$ is achieved from the solutions of the extremum condition
\begin{eqnarray}
m = \tanh\Big[\beta Jm - p\sinh(\beta J_{A})m\Big]\,.
\label{eqstate_an}
\end{eqnarray}
This equation allows one to obtain the critical temperature $\beta_{c}$, which then obeys
\begin{eqnarray}
\beta_{c}J = 1 + p\sinh(\beta_{c}J_{A})\,,
\label{Tc_an}
\end{eqnarray}
and is cleary different from the correspondent expression (\ref{tcq}) from the quenched situation.


\section{Breakdown of the spontaneous magnetization}

This section will analyse the different behaviour of the critical line evaluated in the quenched and annealed approaches, which are
\begin{eqnarray}
\beta_{c}J = 1 + p\tanh(\beta_{c}J_{A}) \qquad\textrm{(quenched case)}
\label{tq}
\end{eqnarray}
and
\begin{eqnarray}
\beta_{c}J = 1 + p\sinh(\beta_{c}J_{A}) \qquad\textrm{(annealed case)}\,.
\label{ta}
\end{eqnarray}

Firstly, it is easy to see that if $p=0$, one recovers the ferromagnetic mean - field result $\beta_{c}J=1$ in both cases, as it should be. Now, let $p$ assume nonzero values. Actually, given $p$ and $J_{A}/J$, the equation (\ref{ta}) may yield two roots, but only the physically reasonable one for the order - disorder transition is chosen. The numerical solutions of the equations are plotted in FIG 2 as a function of $p$. As one can see in the figure, for sufficiently small $J_{A}$ (in the sense that $\beta_{c}J_{A}\ll 1$), the critical temperature describes a line $T_{c}/J \sim 1-p(J_{A}/J)$ in both cases.

\bigskip\bigskip\bigskip

\begin{figure}[htb]
\centering
\includegraphics[width=70.0mm, height=70.0mm]{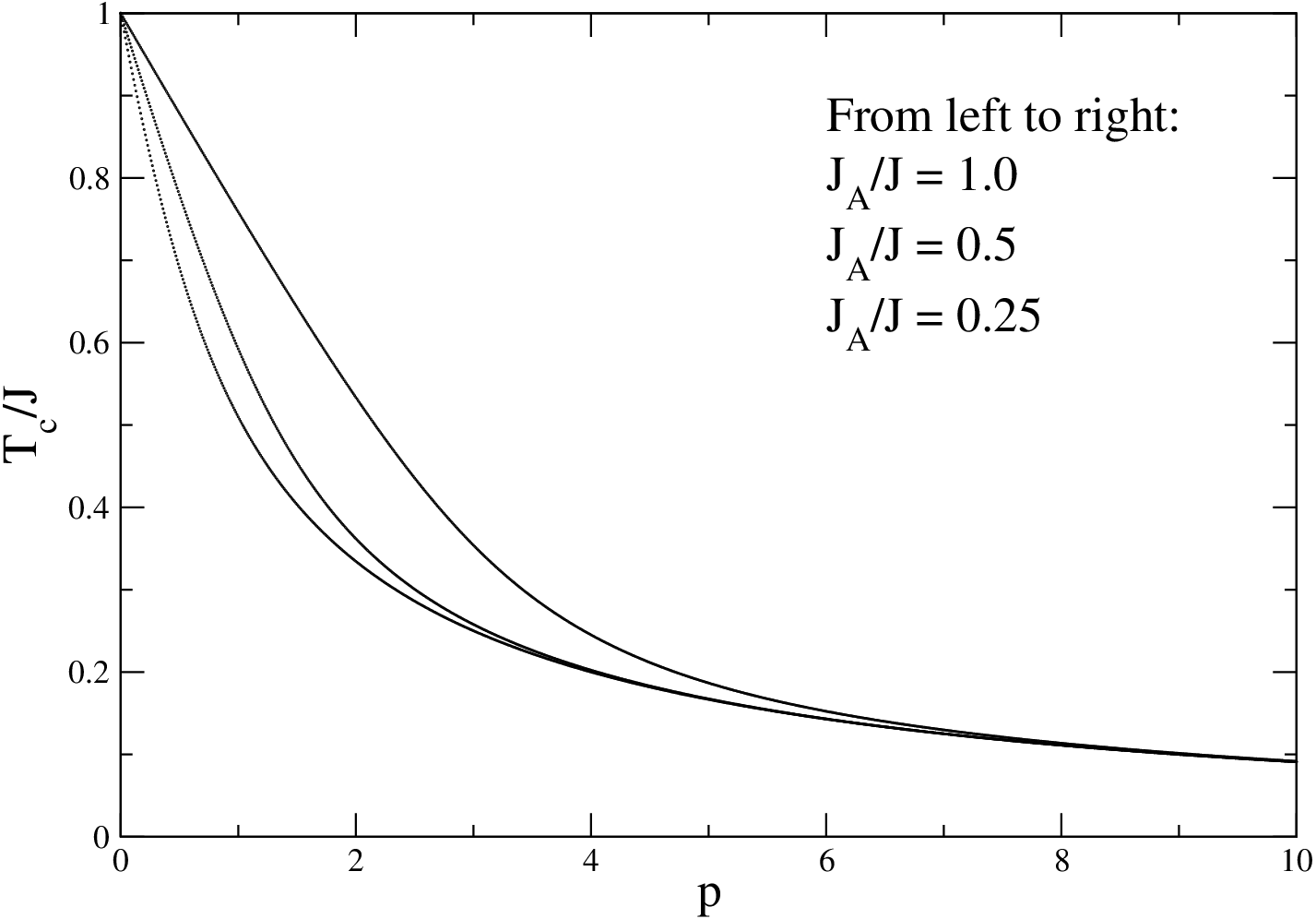}
\hspace{10mm}
\includegraphics[width=70.0mm, height=70.0mm]{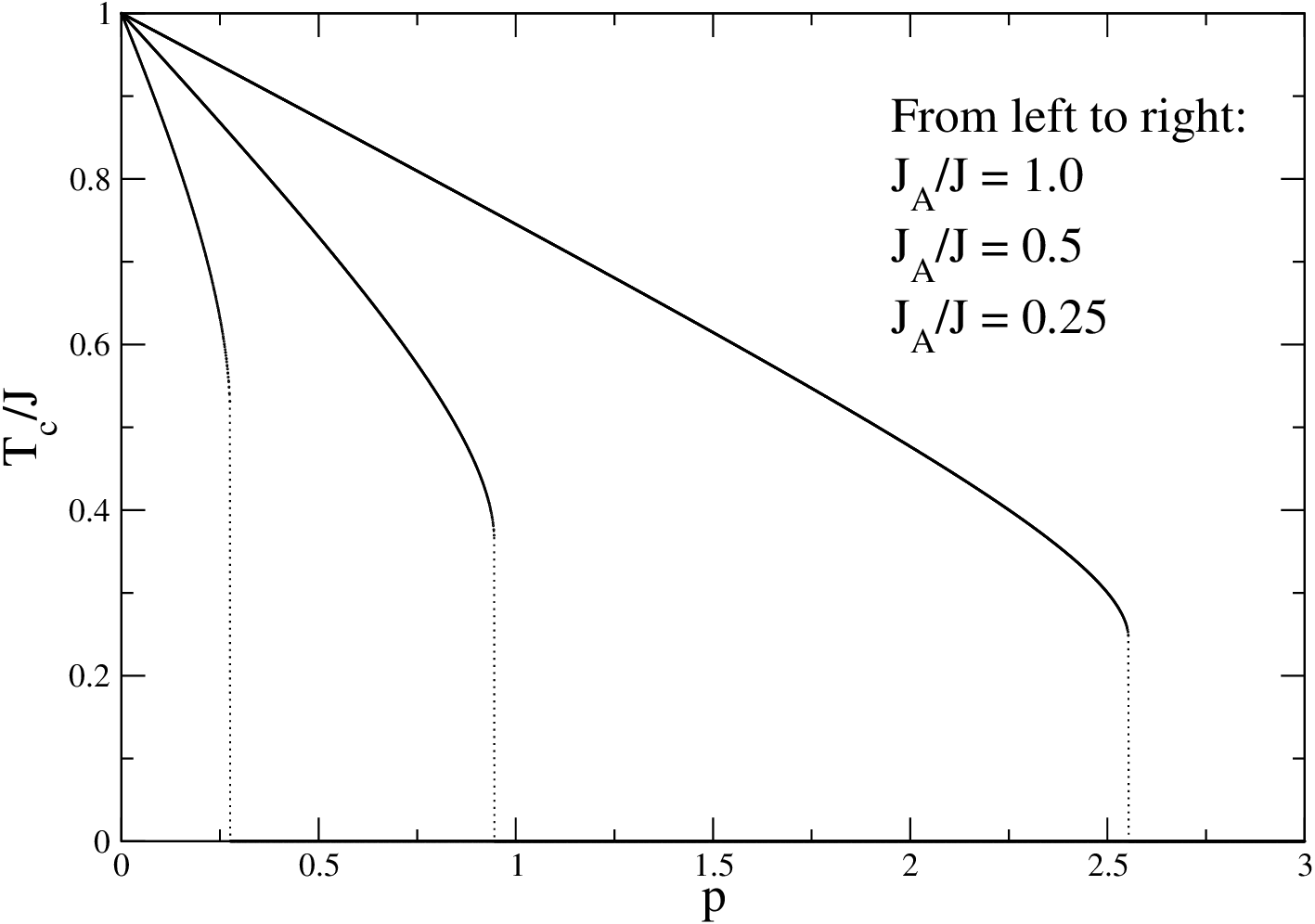}
\bigskip\bigskip
\caption{Dependence of the critical temperature on the mean connectivity $p$ (left: quenched case; right: annealed case).}
\label{2}
\end{figure}

However, for a fixed value of $J_{A}/J$, one can see the difference from the quenched and annealed cases as the mean connectivity $p$ increases. The spontaneous magnetization is broken for sufficiently large values of $p$ in an abrupt way in the annealed case. On the other hand, the critical temperature for the quenched case decays slowly, and reaches $T_{c}=0$ for $p\rightarrow\infty$ only (see (\ref{tq})). One should remember, initially, that each vertex links to another $N-1$ with ferromagnetic interaction $J/N = \mathcal{O}(N^{-1})$ and an antiferromagnetic interaction (which is much stronger in the sense that $J_{A} = \mathcal{O}(1)$) with probability $p/N = \mathcal{O}(N^{-1})$.

In the quenched case, where the configuration $\{a_{xy}\}$ remains frozen during an observational time, only an infinitesimal fraction (of $\mathcal{O}(N^{-1})$) of the links have antiferromagnetic interactions. Suppose that the edge $xy$ (shared between the vertices $x$ and $y$) is one of them; this edge has a ferromagnetic component of intensity $J/N$ and an opposite effect of $J_{A}$ ($\gg J/N$). Although the antiferromagnetic part exceeds the ferromagnetic one, the solely effect is make just the spins $\sigma_{x}$ and $\sigma_{y}$ having opposite signs. Therefore, despite the fact that all the antiferromagnetic interactions dominates over the ferromagnetic interactions in the links where $a_{xy}=1$, the total number of such edges is much smaller than the total number of edges of the network (the thermodynamic limit is taken for a fixed value of $p$), which then becomes predominantely ferromagnetic. This is the cause of persistence of spontaneous magnetization for any finite $p$, as shown in FIG 2 (left).

On the other hand, in the annealing approximation, where the antiferromagnetic links fluctuate during the observation time, one sees an ``averaged'' antiferromagnetic interaction between spins (vertices). This means that, although the mean connectivity $p$ is fixed, the effective number of edges with an antiferromagnetic interaction is much larger (and the effective intensity is also smaller than $J_{A}$). Heuristically speaking, the intensity $J_{A}$ is better distributed over the edges (differently from the quenched case), and the antiferromagnetic effect is better exploited in the annealed case, which makes the magnetization vanishes even for finite values of $p$.


\section{Conclusions}

Throughout this work, the static properties of a diluted antiferromagnet on a ferromagnetic background was examined, with particular emphasis on the breakdown of the spontaneous magnetization of the system. The phase diagram displayed a nonzero magnetization at low temperature regime for any finite mean connectivity $p$ of antiferromagnetic interactions. The disordered phase is constituted by a paramagnetic and a spin - glass phase.

The critical temperature of the order - disorder transition, which indicates the breakdown of the magnetic order of the model, was determined from both the quenched and annealed approaches. The main difference, noticed from FIG 2, relies on the fact that the spontaneous magnetization vanishes for finite values of $p$ in the annealed approximation. This phenomena is observed due to the rapid fluctuation of the random variables $\{a_{xy}\}$, which distributes more efficiently the antiferromagnetic interactions over the whole graph. This means that in the present work, networks are resilient to non - fluctuating random ``attacks'' (even strong ones), while they are weaker to ``annealing attacks'', which turns the antiferromagnetic interaction more accessible to the edges, although weakening its mean strength.


\section*{Acknowledgements}

The authors thank A. V. Goltsev. This work was supported by the project DYSONET.


\end{document}